\documentclass[onecolumn,groupeaddress,showpacs,aps,pra]{revtex4}

\usepackage{latexsym}
\usepackage{amsmath}
\usepackage{amssymb}
\usepackage{latexsym}
\usepackage{color}
\usepackage{graphicx}
\usepackage{bm}
\usepackage{bbm}

\newcommand{\dd}{\mathrm{d}}
\newcommand{\ee}{\mathrm{e}}
\newcommand{\ii}{\mathrm{i}}

\newcommand{\calN}{\mathcal{N}}
\newcommand{\calO}{\mathcal{O}}

\newcommand{\alphadotp}{\vec\alpha\cdot\vec p}

\definecolor{kellygreen}{rgb}{0.3, 0.73, 0.09}
\definecolor{duelferred}{rgb}{0.7, 0.2, 0.1}
\definecolor{garrosgreen}{rgb}{0.1, 0.4, 0.1}
\definecolor{cambridgeblue}{rgb}{0.1, 0.3, 1.0}

\def\half{   {\textstyle{\frac12}}   }

\begin{document}

%
%
\title{Dirac Equations with Confining Potentials}

\author{J. H. Noble}

\affiliation{Department of Physics, Missouri University of Science and
Technology, Rolla, Missouri, MO65409, USA}

\author{U. D. Jentschura}

\affiliation{Department of Physics, Missouri University of Science and
Technology, Rolla, Missouri, MO65409, USA}

\begin{abstract} This paper is devoted to a study of relativistic eigenstates
of Dirac particles which are simultaneously bound by a static Coulomb potential
and added linear confining potentials.  Under certain conditions, despite the
addition of radially symmetric, linear confining potentials, specific
bound-state energies surprisingly preserve their exact Dirac--Coulomb values.
The generality of the ``preservation mechanism'' is investigated.  To this end,
a Foldy--Wouthuysen transformation is used to calculate the corrections to the
spin-orbit coupling induced by the linear confining potentials.  We find that
the matrix elements of the effective operators obtained from the scalar, and
time-like confining potentials mutually cancel for specific ratios of the
prefactors of the effective operators, which must be tailored to the
preservation mechanism.  The result of the Foldy--Wouthuysen transformation is
used to verify that the preservation is restricted (for a given Hamiltonian) to
only one reference state, rather than traceable to a more general relationship
among the obtained effective low-energy operators.  The results derived from
the nonrelativistic effective operators are compared to the fully relativistic
radial Dirac equations. Furthermore, we show that the preservation mechanism
does not affect anti-particle (negative-energy) states.
\end{abstract}


\maketitle

%
%
\section{Introduction}
\label{sec1}

The Dirac-Coulomb Hamiltonian has been the first relativistic bound-state
quantum Hamiltonian ever
investigated~\cite{ItZu1980,SwDr1991a,SwDr1991b,SwDr1991c}, and obvious
generalizations (with confining potentials which describe the binding forces in
nuclei~\cite{HaJeSu1949,GM1949,GM1950I,GM1950II,MK1992,GrMa1996}, 
or for quarks, within the MIT bag model~\cite{ChEtAl1974,MoSa1985})
have phenomenological significance for nuclear and subnuclear physics.  It is
perhaps useful to remember that the investigation of the discrete bound-state
spectrum of the Dirac-Coulomb Hamiltonian has led to the discovery of the
fine-structure of atomic spectra as well as the identification of the relativistic
correction terms that shift the hydrogen spectrum: zitterbewegung term, and
spin-orbit coupling~\cite{BjDr1964,BjDr1965}, while the phenomenological
significance of the spin-orbit coupling for the bound states in nuclei is well
known.  Naturally, one would assume that any added linear confining potentials
(proportional to the radial coordinate) would not only shift the individual
energy levels quantitatively, but even change the spectrum qualitatively,
eliminating the continuous spectrum.  Quite recently, in marked contrast to
this expectation, Franklin and Soares de Castro~\cite{Fr1999,CaFr2000} have
investigated a generalized Dirac Hamiltonian with an added linear confining
potential, or, more precisely, a combination of two added linear confining
potentials, and the surprising conclusion was that specific bound-state
energies are not shifted at all with respect to their Dirac-Coulomb value,
i.e., the added linear confining potential has no influence on the energy
eigenvalue. This observation is very interesting in itself and 
inspires a thorough analysis of the origin, and of the generality
of the ``preservation mechanism''.

In Refs.~\cite{Fr1999,CaFr2000},
the unaffected energies are those of the states with $\kappa=-n$ where $n$ is
the principal quantum number and $\kappa$ is the Dirac angular quantum number.
The papers~\cite{Fr1999,CaFr2000} leave some room for
the interpretation of the generality of the preservation 
mechanism. One might ask whether,
given a specific ratio of the prefactors 
in the confining Hamiltonian, the preservation would
affect only one reference state, whose energy remains unaltered,
or if the parameters could be tailored so that 
the preservation occurs for more than one reference state.
An exact preservation of the Dirac--Coulomb energy 
for a general class of potentials
could otherwise hint at a hidden, exact, hitherto 
insufficiently discussed symmetry of the (generalized) Dirac equation,
with potential phenomenological consequences in related 
areas of physics (nuclear and subnuclear physics).
Our aim here is to analyze the problem 
in detail and to identify the leading relativistic
corrections (effective operators) for the problem studied in
Refs.~\cite{Fr1999,CaFr2000}, via a Foldy--Wouthuysen transformation. 
Expressed differently, our goal is to
investigate whether the results reported in 
Refs.~\cite{Fr1999,CaFr2000} constitute a coincidental relationship,
crucially depending on the fine-tuning of the coupling constants multiplying
the linear confining potential(s), or, if there is a more general physical
pattern behind the energy eigenvalue preservation mechanism.

Our investigation (for a generalization 
see Sec.~\ref{sec5}) focuses on the following generalized Dirac equation,
\begin{equation}
\label{FrankHam}
H \, \psi(\vec r) = E \, \psi(\vec r) \,,
\qquad
H = \vec \alpha \cdot \vec p + \beta \, m- \frac{\lambda}{r} + 
\beta \, \mu \, r + \nu \, r \,,
\end{equation}
where the $\vec \alpha$ and $\beta$ matrices are used in the standard Dirac
representation~\cite{Ro1971a,Ro1971b,SwDr1991a,SwDr1991b,SwDr1991c}, while
$\psi(\vec r)$ is the bispinor (four-component) wave function.  Here, the
binding Coulomb potential corresponds to the term $-\lambda/r$, while the two
added linear confining potentials are given by the terms $\beta \mu r$ and $\nu
\, r$, respectively, and $m$ is the mass of the bound particle.  Surprisingly,
at least one exact solution for a particular eigenstate of Eq.~\eqref{FrankHam}
can be given analytically~\cite{Fr1999}. 
Adopting the conventions of Refs.~\cite{SwDr1991a,SwDr1991b,SwDr1991c},
we assume that the bispinor wave function 
corresponding to the energy eigenstate 
of Eq.~\eqref{FrankHam} has an upper radial component $f$,
and a lower radial component $g$, of the form 
\begin{subequations}
\label{allthestuff}
\begin{align}
\label{allthestuffa}
\psi =& \; \left( \begin{array}{c} f(r) \, \chi_{\kappa_0\,M_0}(\hat r) \\ 
\ii \, g(r) \, \chi_{-\kappa_0\,M_0}(\hat r) \end{array} \right) \,,
\qquad
f(\vec r) = 
\calN \, r^{b-1} \, \exp(-a \,r ) \, \exp(- \tfrac12 \, \alpha^2 \, r^2 ) \,.
\end{align}
Here, $r = |\vec r|$ is the radial coordinate, and 
$\calN$ is a normalization factor, while $\chi_{\kappa\,M}(\hat r) =
\chi_{\kappa_0\,M_0}(\theta, \varphi) $ is the 
spin-angular function~\cite{Ro1971a,Ro1971b}.
The magnetic projection quantum 
number is $M$, the Dirac angular quantum number is 
$\kappa_0 = (-1)^{\ell_0 + j_0 + \tfrac12} \, (j_0+\tfrac12)$
and summarizes both the orbital angular momentum quantum number $\ell_0$
as well as the total angular momentum quantum number $j_0$ 
of the reference state into a single, integer-valued quantum number
(we initially assume that $\kappa_0 = -1$,
which corresponds to an $S$ state).
We use the symbol $M_0$ to denote the magnetic projection
quantum number of the reference state 
because $\mu$ is reserved for the 
coefficient multiplying the scalar confining potential 
in Eq.~\eqref{FrankHam}.
(The zero subscript is reserved for the 
angular quantum numbers of the reference state,
while the corresponding quantum numbers $\ell$, $j$ and $\kappa$
denote a general state; see Sec.~\ref{sec3}.)
Let us assume that the lower component 
$g(r) \, \chi_{-\kappa_0 M_0}(\hat r)$ of the bispinor 
Dirac eigenfunction $\psi$ is given as follows,
\begin{align}
\label{allthestuffb}
g(\vec r) \, \chi_{-\kappa_0 M_0}(\hat r) 
=& \; \gamma (\vec \sigma \cdot \hat r) \, f(r) \, \chi_{\kappa_0 M_0}(\hat r)
\nonumber\\[2ex]
=& \; -\gamma \,
\calN \, r^{b-1} \, \exp(-a \,r ) \, 
\exp\left(- \tfrac12 \, \alpha^2 \, r^2 \right) \,
\chi_{-\kappa_0 M_0}(\hat r) \,.
\end{align}
The vector of Pauli matrices is denoted as $\vec \sigma$,
and $\hat r$ is the position unit vector.
We have used the well-known relation $(\vec \sigma \cdot \hat r) \,
\chi_{\kappa_0 M_0}(\hat r) = - \chi_{-\kappa_0 M_0}(\hat r)$. 
The prefactor $\gamma$ can be shown to simultaneously
fulfill the following relations,
\begin{align}
\label{allthestuffc}
\gamma =& \; \frac{a}{m+E} = \frac{|\kappa_0|-b}{\lambda} 
= \frac{\alpha^2}{\mu-\nu} \,,
\\[0.133ex]
\gamma =& \; \frac{m-E}{a} = \frac{\lambda}{|\kappa_0|+b} 
= \frac{\mu+\nu}{\alpha^2} \,,
\\[0.133ex]
b =& \; \sqrt{1 - \frac{\lambda^2}{\kappa_0^2}} \,,
\quad
a = m \, \frac{\lambda}{|\kappa_0|} \,,
\quad
\alpha^2 = \mu\,\frac\lambda{\kappa_0} \,.
\end{align}
These relations hold provided we assume that 
\begin{equation}
\label{cancel}
\nu = -\mu \, \sqrt{1 - \frac{\lambda^2}{\kappa_0^2}}  \,.
\end{equation}
The normalization constant $\calN$ which ensures that 
$\int \dd^3 r \, |\psi(\vec r)|^2 = 1$, is found to be
\begin{equation}
\calN=\frac1
{\sqrt{2^{-2b}\,a\,b\,\left(1+\gamma^2\right)\,
\alpha^{-2-2b}\,\Gamma\left(2b\right)
U\left(1+b,\frac32,\frac{a^2}{\alpha^2}\right)}} \,,
\end{equation}
where $\Gamma$ denotes  the gamma function and $U$ denotes  Kummer's confluent
hypergeometric function (see Chaps.~6 and~13 of Ref.~\cite{AbSt1972}). 
It has been observed~\cite{Fr1999,CaFr2000}
that the ansatz~\eqref{allthestuffb}
solves the stationary Dirac equation~\eqref{FrankHam} with an 
energy eigenvalue
\begin{equation}
\label{allthestufff}
E = m \, \sqrt{1 - \frac{\lambda^2}{\kappa_0^2}} \,,
\end{equation}
\end{subequations}
which the informed reader will recognize as the
exact Dirac--Coulomb energy for the state with the highest possible
total angular momentum, for a given principle quantum number, namely the
state with $n=n_0=-\kappa_0$.  This is easily verified based on Dirac
theory~\cite{SwDr1991a,SwDr1991b,SwDr1991c}.
In the limit
$\nu \to 0$ and $\mu \to 0$, the Hamiltonian~\eqref{FrankHam} becomes an exact
Dirac--Coulomb Hamiltonian, and the wave function~\eqref{allthestuffb} 
for $\kappa_0 = -1$ tends
toward the ground-state wave function of the Dirac--Coulomb
problem~\cite{SwDr1991a,SwDr1991b,SwDr1991c}.  
Apparently, the addition to two linear confining potentials has
not shifted the ground-state energy at all. 
In (the abstract of) Ref.~\cite{CaFr2000}, the authors state that 
``the method works for the ground state or for the lowest 
orbital state with $\ell = j - 1$, for any $j$''.
This statement may leave some room for interpretation,
in particular, regarding the question of whether the
preservation affects more than one reference state, 
or is limited, for given parameters $n_0$ and $\kappa_0$, 
to one (and only one) reference state.
Furthermore, we should clarify that the state with 
quantum numbers $n = n_0 = -\kappa_0$ actually has 
the highest possible total angular momentum $j_0 = |\kappa_0| - 1/2$ 
and the highest possible orbital angular momentum 
$\ell_0 = j_0 - 1/2$ for given principal quantum number $n_0$.
One would expect that the linear
confinement alters the spectrum not only quantitatively, but even qualitatively
(i.e., it would be assumed to transform the continuous part of the spectrum
into a discrete one, with only bound states). 
Further clarification is definitely required. Throughout the paper, we work in
units with $\hbar = c = \epsilon_0 = 1$.

%
%
\section{Energy Eigenvalue Preservation Mechanism}
\label{sec2}

We aim to investigate the physical mechanism 
at the roots of the intriguing observations reported 
in Eq.~\eqref{allthestuff}, which might hint at a 
hidden symmetry of the (generalized) Dirac equation.
To this end, we carry out a Foldy--Wouthuysen transformation,
which the aim of identifying the 
relevant physical degrees of freedom in the low-energy limit.
The Foldy--Wouthuysen transformation 
is inherently perturbative and therefore tied to a regime 
where the linear confining potentials can still be treated
as perturbations. This, in particular, implies that the 
prefactors of the linear confining potentials have to be
small (parametrically suppressed). 
In the leading approximation, 
use the Schr\"odinger wave equations can be used 
as the unperturbed wave functions.
This is the opposite limit as compared to 
the scenario studied in Ref.~\cite{AbFu1987},
where the authors otherwise consider a situation with 
a confining term that dominates over the rest mass.

Here, we thus investigate the scaling of the 
physical quantities with the perturbative parameter $\lambda$,
a procedure which is 
akin to the well-known so-called $Z\alpha$-expansion in atomic physics.
The following order-of-magnitude estimates are valid for
Schr\"odinger--Pauli bound states~\cite{Pa1993,JePa1996}
\begin{equation}
E = m + \calO(\lambda^2 \, m) \,,
\qquad
|\vec p| = \calO(\lambda \, m) \,,
\qquad
r = \calO\left(\frac{1}{\lambda \, m}\right) \,,
\qquad
\frac{\lambda}{r} = \calO(\lambda^2 \, m) \,.
\end{equation}
A classical analogy is obtained by considering a 
bound nonrelativistic particle on a classical, 
stationary orbit in a $(1/r)$-potential; 
its momentum is of the order $\lambda \, m$.
In order for the confining potentials to represent perturbatively tractable
terms, we therefore have to assume that 
\begin{equation} 
\beta \, \mu \, r = \calO(\lambda^3 \, m) \,,
\qquad
\nu \, r = \calO(\lambda^3 \, m) \,,
\end{equation}
and so the coupling constants of the confining potential must scale at least as 
\begin{equation} 
\mu = \calO(\lambda^4 m^2) \,,
\qquad
\nu = \calO(\lambda^4 m^2) \,.
\end{equation} 
Then, Eq.~\eqref{cancel} implies that, to order $\lambda^4$, we have
\begin{equation} 
\label{cancel2}
\nu = -\mu+\mu \, \frac{\lambda^2}{2\kappa_0^2} + \calO(\lambda^8 m)\,,
\end{equation} 
which is consistent with our previous assumption 
that  the leading-order term in both $\nu$ and $\mu$ 
are of order $\lambda^4 \, m^2$.
Here, we include the correction term of relative order $\lambda^2$ 
which follows by expanding the square root in Eq.~\eqref{cancel}.
With the radial coordinate being of order $1/(\lambda\, m)$,
the matrix elements of the confining potentials 
are of order $\lambda^3 \, m$ and therefore parametrically suppressed
with regard to the main (leading) Schr\"{o}dinger 
energy $-\lambda^2 \, m/(2 n^2)$, where $n$ is the 
principal quantum number. The perturbative hierarchy conveniently orders the 
terms for the application of the Foldy--Wouthuysen (FW)
programs of the  Hamiltonian given in 
Eq.~\eqref{FrankHam}. 

For the FW transformation, one first identifies the
odd part (in bispinor space) of the Hamiltonian in question, with a subsequent
unitary transformation designed to disentangle upper and lower
components. However, in higher orders in the perturbative
parameters, this procedure often tends to introduce higher-order odd terms 
in the resulting Hamiltonian, and therefore it needs to 
be iterated. The process is then
repeated until all odd parts of the Hamiltonian are eliminated to the desired
order in perturbation theory~\cite{FoWu1950,BjDr1964,BjDr1965}. 
In our case, we will keep all terms up to
order $\lambda^5 \, m$ in the energy. We identify the
odd part of Eq.~\eqref{FrankHam} as
\begin{equation}
\calO=\alphadotp\,,\qquad 
S=-\ii\frac{\beta\,\calO}{2m}\,,\qquad 
U=\ee^{-\ii S}\,,
\end{equation}
where we have also constructed the Hermitian operator $S$ and the 
unitary rotation operator $U$.  The rotation is then given 
as~\cite{FoWu1950,BjDr1964}
\begin{equation}
H'=U\,H\,U^+\approx H+\ii[S,H]+\frac{(\ii)^2}{2!}[S,[S,H]]
+\frac{(\ii)^3}{3!}[S,[S,[S,H]]]
+\dots\,.
\end{equation}
We now apply the FW transformation and find
\begin{align}
H'=&\,\beta\left(m+\frac{{\vec p}^{\,2}}{2m}-\frac{{\vec p}^{\,4}}{8m^3}\right)
-\frac\lambda r+\frac1{8m^2}
\left[\alphadotp,\left[\alphadotp,\frac\lambda r\right]\right]\\
&\,+\beta\,\mu\left(r-\frac{\{\alphadotp,\{\alphadotp,r\}\}}{8m^2}\right)
+\nu\left(r-\frac{[\alphadotp,[\alphadotp,r]]}{8m^2}\right)+\calO'\,,\nonumber
\end{align}
where
\begin{align}
\calO'=&\,-\frac{{\vec p}^{\,2}\alphadotp}{3m^2}+\frac{{\vec p}^{\,4}\alphadotp}{30m^4}
-\frac\beta{2m}\left[\alphadotp,\frac\lambda r\right]-
\frac{\mu}{2m}\{\alphadotp,r\}
+\frac{\beta\,\nu}{2m}[\alphadotp,r]\\
&\,+\frac\beta{48m^3}\left[\alphadotp,
\left[\alphadotp,\left[\alphadotp,\frac\lambda r \right]\right]\right]\,.\nonumber
\end{align}
Note that while $\calO\sim\lambda m$, $\calO'$ is of higher order,
namely $\calO'\sim\lambda^3m$.
In order to proceed, we must perform another iteration of the
FW transformation, with 
\begin{equation}
S'=-\ii\frac{\beta\,\calO'}{2m}\,,\qquad
U'=\ee^{-\ii S'}\,,
\end{equation}
which yields
\begin{align}
H''=&\,\beta\left(m+\frac{{\vec p}^{\,2}}{2m}-\frac{{\vec p}^{\,4}}{8m^3}\right)
-\frac\lambda r+\frac1{8m^2}
\left[\alphadotp,\left[\alphadotp,\frac\lambda r\right]\right]\\
&\,+\beta\,\mu\left(r-
\frac{\{\alphadotp,\{\alphadotp,r\}\}}{8m^2}\right)
+\nu\left(r -
\frac{[\alphadotp,[\alphadotp,r]]}{8m^2}\right)+\calO''\,,\nonumber
\end{align}
where
\begin{equation}
\calO''=\frac{{\vec p}^{\,4}\alphadotp}{6m^4}
+\frac\beta{6m^3}\left[{\vec p}^{\,2}\alphadotp,\frac\lambda r\right]
+\frac\beta{8m^3}\left\{{\vec p}^{\,2},\left[\alphadotp,\frac\lambda r\right]\right\}
+\frac1{4m^2}
\left[\left[\alphadotp,\frac\lambda r\right],\frac\lambda r\right]\,.
\end{equation}
The new odd part is of order $\lambda^5m$, while
to the same order, the even part remains the same.
For the third transformation, we use
\begin{equation}
S''=-\ii\frac{\beta\,\calO''}{2m}\,,\qquad
U''=\ee^{-\ii S''}\,,
\end{equation}
and perform a final iteration of the FW transformation, yielding
\begin{align}
H^{(FW)}=&\,\beta\left(m+\frac{{\vec p}^{\,2}}{2m}-\frac{{\vec p}^{\,4}}{8m^3}
\right)
-\frac\lambda r+\frac1{8m^2}
\left[\alphadotp,\left[\alphadotp,\frac\lambda r\right]\right]\\
&\,+\beta\,\mu\left(r-
\frac{\{\alphadotp,\{\alphadotp,r\}\}}{8m^2}\right)
+\nu\left(r-\frac{[\alphadotp,[\alphadotp,r]]}{8m^2}\right)\,,\nonumber
\end{align}
which will be rewritten with the help of the identities
\begin{subequations}
\begin{align}
\left[\alphadotp,\left[\alphadotp,\frac1r\right]\right]
=&\,4\pi\delta^{(3)}(\vec r)+2\frac{\vec\Sigma\cdot\vec L}{r^3}\,,
\\[0.133ex]
\{\alphadotp,\{\alphadotp,r\}\}=& \; 2\{\vec p^{\,2}, r\}+2 \beta \frac{K}r\,,
\\[0.133ex]
[\alphadotp,[\alphadotp,r]]=&\; -2 \beta \, \frac{K}r\,,
\\[0.133ex]
K \equiv & \; \beta \, ( \vec\Sigma\cdot\vec L + 1) = \left(
\begin{array}{cc}
\vec\sigma\cdot\vec L + 1 & 0\\ 
0 & - (\vec\sigma\cdot\vec L + 1) \end{array}\right) \,.
\end{align}
\end{subequations}
While the $(4\times 4)$-bispinor operator $K$ 
commutes with the Hamiltonian, $[ K, H ] = 0$,
the $(2 \times 2)$-submatrices of the operator $K$ 
have the properties~\cite{Ro1961},
\begin{equation}
\label{properties}
( \vec\sigma\cdot\vec L + 1 ) \,
\chi_{\kappa_0 M_0} = -\kappa_0\,\chi_{\kappa_0 M_0}\,,
\qquad
\kappa_0 = (-1)^{j_0 + \ell_0 + 1/2} \, \left( j_0 + \tfrac12 \right) \,.
\end{equation}
The orbital angular momentum is obtained as 
$\ell_0 = | \kappa_0 + 1/2 | - 1/2$, and the 
total angular momentum reads $j_0 = | \kappa_0 | - 1/2$.
The eigenfunctions of the Hamiltonian given in 
Eq.~\eqref{allthestuffa} are also eigenfunctions
of $K$ with eigenvalue $-\kappa_0$. The $(4\times 4)$ FW-transformed
Hamiltonian finally reads as follows,
\begin{align}
\label{almost}
H^{(FW)}=&\,\beta\left(m+\frac{{\vec p}^{\,2}}{2m}
-\frac{{\vec p}^{\,4}}{8m^3}\right)
-\frac\lambda r+\frac{\pi\lambda}{2m^2}\delta^{(3)}(\vec r)
+\frac\lambda{2m^2r^3}\vec\Sigma\cdot\vec L\\
&\,+(\nu+\beta\,\mu)r+(\beta\,\nu-\mu)\frac{K}{4m^2r}
-\beta\,\mu\frac{\{{\vec p}^{\,2},r\}}{4m^2}\,.\nonumber
\end{align}
One now uses Eq.~\eqref{cancel2} and
expresses all terms in terms of $\mu$, 
keeping all terms in the the Hamiltonian 
up to order $\lambda^5 \, m$. This results in
\begin{subequations}
\label{FWHam}
\begin{align}
H^{(FW)}=&\, H_{DC} + H_C \,,\\
\label{FWHamDC}
H_{DC} =& \; 
\beta m
\underbrace{ + \beta \frac{{\vec p}^{\,2}}{2m} -\frac\lambda r }_%
{\mbox{$\calO(\lambda^2 \, m)$}}
\underbrace{- \beta \frac{{\vec p}^{\,4}}{8m^3}
+\frac{\pi\lambda}{2 \, m^2}\delta^{(3)}(\vec r)
+\frac\lambda{2 \, m^2 \, r^3}\vec\Sigma\cdot\vec L }_%
{\mbox{$\calO(\lambda^4 \, m)$}} \,, \\[2ex]
\label{FWHamC}
H_{C} =& \; 
\underbrace{\mu \left( \beta-1 \right) \, r}_{\mbox{$\calO(\lambda^3 \, m)$}}
+ \underbrace{ \frac{\lambda^2 \, \mu \, r}{2\kappa_0^2} 
- (\beta+1) \, \mu \, \frac{\vec\Sigma \cdot \vec L + 1}{4 \, m^2 \, r}
-\beta\,\mu \, \frac{\{ \vec p^{\,2},r\}}{4m^2} }_%
{\mbox{$\calO(\lambda^5 \, m)$}}\,,
\end{align}
\end{subequations}
where $H_{DC}$ is the FW transformed Dirac-Coulomb
Hamiltonian~\cite{JeNo2014jpa,BjDr1964}, and $H_C$ is a perturbative term which
was induced by the addition of the confining potentials. We see that the
perturbation in the original Hamiltonian, under the conditions imposed in
Eq.~\eqref{cancel} with the approximation made in Eq.~\eqref{cancel2}, results
in the sum of three additional terms of order $\lambda^5 m$ in the FW
transformed Hamiltonian, beyond the leading confining term of order $\lambda^3
\,m$, which is unaffected by the FW program in comparison to
Eq.~\eqref{FrankHam}. The first of these terms simply is a linear
potential.  It is followed by a spin-orbit coupling term, proportional to
$\vec\Sigma \cdot \vec L/r$ instead of $\vec\Sigma \cdot \vec L/r^3$ as in
$H_{DC}$.  The former is the result of the FW transformation of a linear
potential while the latter comes from the transformation of a $1/r$ potential.
Finally we see that the linear perturbation also induces a kinetic correction
term to the particle's orbit, whose functional form is akin to the magnetic
exchange term in the Breit Hamiltonian~\cite{BeLiPi1982vol4}.

According to Eq.~\eqref{FWHamC},
our calculation shows a very intriguing ``duality'' 
of the cancellation of the leading confining term 
for a particle state [term proportional to 
$\mu \left( \beta-1 \right) \, r \to \beta - 1\to 0$ for the 
upper bispinor], while the spin-orbit term 
[proportional to $(\beta+1) \, (\vec\Sigma \cdot \vec L + 1)
\to \beta + 1 \to 0$ vanishes for antiparticle states.
Furthermore, for particle states, the spin-orbit coupling term has the 
``opposite'' sign in the prefactor of the 
term $\vec\Sigma \cdot \vec L$ as compared to the Dirac--Coulomb problem,
where it is otherwise known to be proportional to
$+\vec\Sigma \cdot \vec L/(m^2 \, r^3)$.

For a particle [denoted with a superscript ``(+)''] as opposed to an 
antiparticle state, we may replace $\beta\rightarrow1$ 
and isolate the upper $2\times2$ submatrix of
the Hamiltonian,
\begin{subequations}
\label{FWHamplus}
\begin{align}
\label{FWHamDCplus}
H^{(+)}_{DC} =& \; 
m \underbrace{ + \frac{{\vec p}^{\,2}}{2m} -\frac\lambda r }_%
{\mbox{$\calO(\lambda^2 \, m)$}}
\underbrace{- \frac{{\vec p}^{\,4}}{8 \, m^3}
+\frac{\pi\lambda}{2 \, m^2} \, \delta^{(3)}(\vec r)
+\frac\lambda{2 \, m^2 \, r^3} \, \vec\sigma\cdot\vec L }_%
{\mbox{$\calO(\lambda^4 \, m)$}} \,, 
\\[0.133ex]
\label{FWHamCplus}
H_C^{(+)} =& \; \underbrace{\mu\left(\frac{\lambda^2}{2\kappa_0^2} r -
\frac{\vec\sigma \cdot \vec L + 1}{2m^2r}
-\frac{\{{\vec p}^{\,2},r\}}{4m^2}\right)}_{\mbox{$\calO(\lambda^5 \, m)$}} \,.
\end{align}
\end{subequations}
After the FW transformation, the linear confining 
potential $\beta \mu r $ from Eq.~\eqref{FrankHam}
remains confining for particle states, while the 
potential $\nu \, r$, given the relations~\eqref{cancel} and~\eqref{cancel2},
transforms into an anti-confining potential, 
leading to conceivable cancellation.
The explicit appearance of the reference state quantum number $\kappa_0$ 
in Eq.~\eqref{FWHamCplus} reminds us of the fact that the 
relation~\eqref{cancel} depends on the reference state.

We now recall some basic facts about the Dirac angular
momentum quantum number $\kappa_0$.
The possible values for $j_0$ are $j_0=\ell\pm\half$, and
it is instructive to distinguish
the cases of negative and positive $\kappa_0$, expressing $\kappa_0$
in terms of $\ell_0$, 
\begin{equation}
\label{KappaEll}
\kappa_0
= \left\{\begin{array}{cc} -(j_0+\tfrac12) & \quad j_0 = \ell_0 + \tfrac12 \\[2ex]
+(j_0+ \tfrac12) & \quad j_0 = \ell_0 - \tfrac12 \end{array}\right.
\quad = \quad \left\{\begin{array}{cc}-(\ell_0+1) & 
\qquad j_0 = \ell_0 + \tfrac12 \\[2ex]
\ell_0 & \qquad j_0 = \ell_0 - \tfrac12 \end{array}\right.\,.
\end{equation}
The relation $\kappa_0(\kappa_0+1) =\ell_0 \, (\ell_0 + 1)$ holds regardless of 
the relative orientation of the orbital angular momentum $\ell_0$ and 
spin projection $\pm \tfrac12$. 
Using the quantum numbers $n_0$ and $\kappa_0$,
as well as the magnetic momentum projection $M$,
one may describe a state which otherwise needs the 
quantum numbers $n_0$, $\ell_0$, and $j_0$.
In the usual $n_0 \ell_j$ notation, examples are
as follows. For the $1S_{1/2}$ ground state,
we have $n_0=1$, $\kappa_0=-1$,
while the $2P$ states are given as follows.
Namely, for $2P_{1/2}$, we have $n_0=2$, $\kappa_0=1$,
while for $2P_{3/2}$, we have $n_0=2$, $\kappa_0=-2$.
The $(n_0=3)$ state are $3D_{3/2}$,
which $n_0=3$ and $\kappa_0=2$,
and $3D_{5/2}$, with $n_0=3$ and $\kappa_0=-3$.
The $P$ states with $n_0=2$ are 
$3P_{1/2}$ with $\kappa_0=1$, and
$3P_{3/2}$ with $\kappa_0=-2$.
For given $n_0$, $\kappa_0$ attains all 
integer values in the interval $(-n_0, -n_0+1, \ldots, n_0-1)$,
excluding zero.

Let us now look at the expectation value of the perturbation for a
$\left|n_0\,\kappa_0\right>$ reference state. First we note that in such a
state, the expectation value of $\vec\sigma\cdot\vec L$ is just 
$\kappa_0-1$. We can use
the Schr\"{o}dinger-Pauli states~\cite{Ro1971a,Ro1971b} as basis states when
evaluating the energy perturbation. For a nonrelativistic 
wave function of the form
$\phi(\vec r)=R_{n_0\,\ell_0}(r) \chi_{\kappa_0\,M_0}(\hat r)$, 
the radial part $R_{n_0\,\ell_0}(r)$ is given by the radial solution to the
Coulomb--Schr\"{o}dinger Hamiltonian $\left(\frac{\vec
p^{\,2}}{2m}-\frac\lambda r\right)$. In leading order,
using formulas from Ref.~\cite{BeSa1957}, we then find that
\begin{subequations}
\label{matelem}
\begin{align}
\label{exp}
E_{n_0 \, \kappa_0} \approx & \;
\left< n_0 \kappa_0 \left| m + \frac{\vec p^{\,2}}{2m} - \frac{\lambda}{r} 
\right| n_0 \kappa_0 \right> =
m - \frac{\lambda^2 \, m}{2 \, n_0^2}\,,
\\[0.133ex]
\langle r \rangle_0 =& \; 
\left< n_0 \kappa_0 \left| r \right| n \kappa_0 \right> =
\frac{3n_0^2-\kappa_0(\kappa_0+1)}{2\lambda m} \,,
\\[0.133ex]
\left< r^{-1} \right>_0 =& \;
\left< n_0 \kappa_0 \left| r^{-1} \right| n \kappa_0 \right> =
\frac{\lambda \, m}{n_0^2}\,.
\end{align}
\end{subequations}
Independent of the magnetic projection $M_0$ of the reference state,
and with the help of Eq.~\eqref{properties}, one then finds that 
\begin{align}
\label{R2}
\left< n_0 \kappa_0 \left| H_{C}^{(+)} \right| n_0 \kappa_0 \right>
=& \; \frac{\mu \lambda}{4 m}
\frac{(n_0-\kappa_0)\,(n_0+\kappa_0)\,(3n_0^2+\kappa_0 (\kappa_0 -1))}%
{\kappa_0^2 \, n_0^2} = 0 \,,
\end{align}
because we explicitly assume that $n_0 = -\kappa_0$.
As anticipated, due to the fact that the energy is equivalent to that of the
corresponding state of the Dirac--Coulomb 
Hamiltonian, we see that the perturbation
disappears, but only because we have adjusted 
the relationship~\eqref{cancel2} among the prefactors of the 
two confining potentials.
One might ask if this preservation is a more general
phenomenon, or specific to the particular reference state at hand.

%
%
\section{Confining Terms for General Positive--Energy States}
\label{sec3}

Let us consider the perturbed particle Hamiltonian, $H_C^{(+)}$,
given in Eq.~\eqref{FWHamCplus},
and its expectation value for a general reference
state $|n \, \kappa \rangle \neq |n_0 \, \kappa_0 \rangle$.
Again, using formulas from Ref.~\cite{BeSa1957},
we find that the first-order energy shift $\Delta E_{n\kappa}$ 
reads as 
\begin{equation}
\label{corrterm}
\Delta E_{n \kappa} =
\left< n \kappa \left| H_{C}^{(+)} \right| n \kappa \right> =
\frac{\mu\lambda}{4 m} \, \left(
\frac{3 n^2-\kappa(\kappa+1)}{\kappa_0^2}
- \frac{n^2+\kappa(\kappa-1)}{n^2}\right)\,.
\end{equation}
The question then is whether at least the leading correction 
to the energy due to the confining potential, given by 
Eq.~\eqref{corrterm}, vanishes for states other than 
$|n \, \kappa \rangle = |n_0 \, \kappa_0 \rangle$.
We recall that $\Delta E_{n \kappa}$ is 
of order $\lambda^5$ as $\mu$ is of order $\lambda^4$. 
Now, $\Delta E_{n\kappa}$ vanishes provided
\begin{equation}
\label{RHS}
\kappa_0^2 = n^2 \, \Xi^2 \,,
\qquad
\Xi^2 = \frac{3n^2-\kappa(\kappa+1)}{n^2+\kappa(\kappa-1)} \,.
\end{equation}
If we set $\kappa = \pm n$, then $\Xi$ evaluates to unity, 
so that $\kappa_0 = \pm n$. The only sign which can be realized
for bound states pertains to $n = -\kappa$
[see the text following Eq.~\eqref{KappaEll}],
in which case $\kappa = \kappa_0 = -n = -n_0$, 
and we reproduce the case already discussed in Eq.~\eqref{R2}.

In order to check whether the perturbation vanishes for any other state,
one investigates if the case $\kappa_0^2 = n^2 \, N^2$ may 
occur, where $N$ is an integer greater than unity, i.e., if
the case $\Xi^2 = N^2$ with $N \geq 2$ can be realized.
We set 
\begin{equation}
\label{FracN}
\frac{3n^2-\kappa(\kappa+1)}{n^2+\kappa(\kappa-1)}=N^2\,.
\end{equation}
Again, if we set $N=1$, then we find $\kappa^2=n^2$, which is 
the physically relevant solution already discussed. 
For $N>1$, we rearrange Eq.~\eqref{FracN} to find
\begin{equation}
\label{FalseEq}
(N^2-3)n^2 = \kappa((N^2-1)-(N^2+1)\kappa)\,.
\end{equation}
Under the assumption that $N \geq 2$, one ascertains that
\begin{equation}
\label{ahahahaaaa}
(N^2-3)n^2>0\,,\qquad
\kappa((N^2-1)-(N^2+1)\kappa)<0\,,
\end{equation}
regardless of whether $\kappa$ is negative or positive.  Thus
Eq.~\eqref{FalseEq} {\em cannot} be fulfilled, and the only possible
integer value for $N^2$ is $1$.  It then follows that $\Delta E_{n \kappa}=0$
{\em if and only if} $\kappa=\kappa_0=-n=-n_0$.

Summarizing, we can tailor the prefactors of the two confining
potentials, according to the modified relation~\eqref{cancel},
\begin{equation}
\nu = -\mu \, \sqrt{1 - \frac{\lambda^2}{\kappa_0^2}}  \,,
\end{equation}
and obtain the preservation for the state $|n_0 \kappa_0 \rangle$
with $n_0 = -\kappa_0$, which happens to be 
the state of maximum total angular momentum for given 
principal quantum number. However, once $\kappa_0$ is fixed,
no other reference states are affected by the preservation,
and in fact, are shifted with regard to their 
unperturbed (Dirac--Coulomb) values.
This constitutes a restriction to the preservation 
mechanism for the energy eigenvalue. 

%
%
\section{Effective Confining Potentials for Antiparticle States}
\label{sec4}

Let us recall Eq.~\eqref{FWHam} and discuss how it pertains 
to antiparticle states,
\begin{subequations}
\label{FWHam2}
\begin{align}
H^{(FW)}=&\, H_{DC} + H_C \,,\\
H_{DC} =& \;
\beta m + \beta \frac{{\vec p}^{\,2}}{2m} -\frac\lambda r 
- \beta \frac{{\vec p}^{\,4}}{8m^3}
+\frac{\pi\lambda}{2 \, m^2}\delta^{(3)}(\vec r)
+\frac\lambda{2 \, m^2 \, r^3}\vec\Sigma\cdot\vec L \,, \\[2ex]
H_{C} =& \;
\mu \left( \beta-1 \right) \, r
+ \frac{\lambda^2 \, \mu \, r}{2 \, \kappa_0^2}
- (\beta+1) \, \mu \, \frac{\vec\Sigma \cdot \vec L + 1}{4 \, m^2 \, r}
-\beta\,\mu \, \frac{\{ \vec p^{\,2},r\}}{4m^2} \,.
\end{align}
\end{subequations}
The FW transformation has disentangled the particle from the 
antiparticle degrees of freedom, as described by the Dirac $\beta$ 
matrix, 
\begin{equation}
\beta = \left( \begin{array}{cc} \mathbbm{1}_{2 \times 2} & 0 \\
0 &  -\mathbbm{1}_{2 \times 2} \end{array} \right) \,.
\end{equation}
Formally, the Hamiltonian $H^{(FW)}$, being the time evolution 
operator, has negative energy eigenvalues corresponding to the 
antiparticle states; however, the reinterpretation 
principle~\cite{ItZu1980} dictates that the physical energy
operator for antiparticles equals the negative of the 
lower $(2 \times 2)$ submatrix of the Hamiltonian~\eqref{FWHam2}.
We endow this physical Hamilton operator for the 
antiparticle states with the superscript ``$(-)$'' and write
\begin{subequations}
\label{HminusDC}
\begin{align}
\label{HminusDCa}
H^{(-)}_{DC} =& \; 
m \underbrace{ + \frac{{\vec p}^{\,2}}{2m} + \frac\lambda r }_%
{\mbox{$\calO(\lambda^2 \, m)$}}
\underbrace{- \frac{{\vec p}^{\,4}}{8 \, m^3}
- \frac{\pi\lambda}{2 \, m^2} \, \delta^{(3)}(\vec r)
- \frac\lambda{2 \, m^2 \, r^3} \, \vec\sigma\cdot\vec L }_%
{\mbox{$\calO(\lambda^4 \, m)$}} \,, 
\\[0.133ex]
\label{HminusDCb}
H_C^{(-)} =& \; 
\underbrace{2 \mu \, r}_{\mbox{$\calO(\lambda^3 \, m)$}} 
\underbrace{ - \mu \, \frac{\lambda^2}{2 \, \kappa_0^2}r - 
\mu \, \frac{\{{\vec p}^{\,2},r\}}{4m^2}}_%
{\mbox{$\calO(\lambda^5 \, m)$}} \,.
\end{align}
\end{subequations}
Here, the order-of-magnitude estimates in regard to $\lambda$
pertain to the expectation values that would otherwise be obtained 
for bound Schr\"{o}dinger--Pauli reference states. 
However, as is well known~\cite{Ro1971a,Ro1971b}
and manifest in Eq.~\eqref{HminusDC}, for antiparticles,
the terms of order $\calO(\lambda^2 \, m)$ in 
Eq.~\eqref{HminusDC} actually describe 
a repulsive (``positron'') Schr\"{o}dinger Hamiltonian
\begin{equation}
H^{(-)}_S = \frac{{\vec p}^{\,2}}{2m} + \frac\lambda r
\end{equation}
where we ignore the rest mass term from Eq.~\eqref{HminusDCa}.
The spectrum of $H^{(-)}_S$ consists only of continuum states,
reflecting the ``sign change'' of the Coulomb potential $\lambda/r$ 
for antiparticles. This means that the analysis outlined 
above for bound particle states is not applicable to the 
antiparticle states in the FW transformed Hamiltonian.
Furthermore, the two confining potentials, in sharp contrast to 
the preservations observed for 
the particle Hamiltonian, add up for antiparticles
to a confining linear term $2\mu \, r$
which, perturbatively, is of order $\lambda^3 \, m$.

As already stated, the order-of-magnitude estimates given in powers of
$\lambda$ in Eq.~\eqref{HminusDC} are relevant to bound states and otherwise
are consistent with the classical analogue of bound particles orbiting the
binding $(-\lambda/r)$-potential with velocities of order $\lambda \, c$.  For
antiparticles, in the repulsive $(+\lambda/r)$-potential, the physics changes.
Here, one can typically assume that the energy of the particle with regard to
the continuum threshold in the repulsive potential is large compared to the
binding energy scale of the sign-reversed potential.  This means, in
particular, that the wave functions of the continuum states are typically
spread farther out than the scale $\langle r \rangle \sim 1/(\lambda\,m)$,
which is typical of bound states.
For states with $\langle r \rangle \gg 1/(\lambda\,m)$,
the Coulomb potential is suppressed in comparison to the 
effective confining term $2 \mu \, r$ given in Eq.~\eqref{HminusDCb}.
In this regime, we can thus approximate
\begin{equation}
\label{HminusEFF}
H^{(-)} = H^{(-)}_{DC} +  H^{(-)}_{C} 
\approx m + \frac{{\vec p}^{\,2}}{2m} - \frac{{\vec p}^{\,4}}{8 \, m^3}
+ 2 \mu \, r \,,
\end{equation}
and treat the repulsive part of the Coulomb potential,
given by the term $\lambda/r$, as a perturbation.
As compared to the particle states, the effective 
Hamiltonian for antiparticle states, obtained after the 
FW transformation, leads up to no compensation between the 
two confining potentials, which add up rather than cancel.
This observation illustrates the different physics that 
result from one and the same generalized Dirac Hamiltonian,
given in Eq.~\eqref{FWHam}, in the case where 
the Dirac Hamiltonian is not charge conjugation invariant
and therefore may result in fundamentally different
physics for particles and corresponding antiparticles.

The confining potential $2\mu \, r$ leads to 
a transformation of the spectrum in terms of a discrete spectrum.
For not--too--large value of the coupling parameter $\mu$, 
this spectrum will contain a series of relatively 
dense eigenvalues.

%
%
\section{Fully Relativistic Radial Equation}
\label{sec5}

Up to this point, we have employed a Foldy--Wouthuysen transformation 
in order to disentangle the upper and lower components of the Hamiltonian 
and wave function, i.e., the positive- and negative-energy states.
It is useful to contrast this discussion with the radial wave 
equations that are obtained from the fully relativistic formalism.
We assume that, without the added linear confining 
potentials, the initial Dirac equation is of the form
\begin{equation}
\label{A1}
\left(\vec\alpha\cdot\vec p+\beta\,m+
V_0(r)\right)\psi(\vec r)= E\psi_0(\vec r)\,,\qquad
\psi_0(\vec r)=\left(\begin{array}{c}
f_0(r)\chi_{\kappa_0\,M_0}\\
\ii g_0(r)\chi_{-\kappa_0\,M_0}
\end{array}\right)\,,
\end{equation}
where we additionally assume that the 
``unperturbed'' solution $\psi_0(\vec r)$ while $V_0(r)$ 
denotes some sort of ``Coulomb potential''. 
We then add a linear combination of radially symmetric
potentials $V_1$ and $V_2$ 
to the equation that defines the energy eigenvalue $E$,
\begin{subequations}
\label{A2}
\begin{align}
& \; \left(\vec\alpha\cdot\vec p+\beta\,m+
V_0(r)+\beta\,V_1(r)+V_2(r)\right)\psi(\vec r)=E\psi(\vec r) \,,
\\[2ex]
& \; \psi(\vec r)= \left(\begin{array}{c}
f(r) \, \chi_{\kappa_0\,M_0}\\
\ii g(r) \, \chi_{-\kappa_0\,M_0}
\end{array}\right)\,,
\end{align}
\end{subequations}
and investigate under which conditions
we can find radial wave functions $f(r)$ and $g(r)$
[which may be different from the $f_0(r)$ and $g_0(r)$]
while the energy eigenvalue $E$ is 
preserved in Eqs.~\eqref{A1} and~\eqref{A2}. 
The radial equations for
the original ``unperturbed'' Dirac--Coulomb Hamiltonian are then
\begin{subequations}
\begin{align}
\left(\frac{\partial}{\partial r}+\frac{\kappa_0+1}r\right)f_0(r)
=&\,\left(E+m+V_0(r)\right)g_0(r)\,,\\
\left(-\frac{\partial}{\partial r} +\frac{\kappa_0-1}r\right)g_0(r)
=&\,\left(E-m+V_0(r)\right)f_0(r)\,,
\end{align}
\end{subequations}
while the radial equations for the perturbed Hamiltonian read as
\begin{subequations}
\begin{align}
\left(\frac{\partial}{\partial r}+\frac{\kappa_0+1}r\right)f(r)
=&\,\left(E+m+V_0(r)+V_1(r)-V_2(r)\right)g(r)\,,\\
\label{PertHamRadII}
\left(-\frac{\partial}{\partial r} +\frac{\kappa_0-1}r\right)g(r)
=&\,\left(E-m+V_0(r)-V_1(r)-V_2(r)\right)f(r)\,.
\end{align}
\end{subequations}
For a given energy eigenvalue $E$,
one may rescale of the ``unperturbed'' radial wave functions as follows,
\begin{equation}
\label{PertORel}
f(r)=\ee^{h(r)} \, f_0(r)\,,\qquad
g(r)=\ee^{h(r)} \, g_0(r)\,,\qquad
\psi(\vec r)=\ee^{h(r)} \, \psi_0(\vec r)\,.
\end{equation}
This implies that
\begin{subequations}
\label{PartialRelation}
\begin{align}
\frac{\partial h(r)}{\partial r}f(r)=&\,(V_1(r)-V_2(r)) \, g(r)\,,\\
\frac{\partial h(r)}{\partial r}g(r)=&\,(V_1(r)+V_2(r)) \, f(r)\,.
\end{align}
\end{subequations}
Multiplying both sides of these equations, we find
the condition
\begin{equation}
\label{PartialH}
\frac{\partial h(r)}{\partial r} 
= \pm\sqrt{\left(V_1(r)\right)^2-\left(V_2(r)\right)^2}\,.
\end{equation}
Based on these considerations, one might conclude that it is
possible to add any combinations of radial potentials $V_1$ and 
$V_2$ to a solvable Dirac Hamiltonian with 
potential $V_0(r)$, provided that the added potentials 
$V_1(r)$ and $V_2(r)$ fulfill the condition that
$\left(V_1(r)\right)^2\ge\left(V_2(r)\right)^2$, and quickly find a solution 
using Eq.~\eqref{PertORel}.
The state $\psi_0(\vec r)$ has retained its energy $E$ even after
the addition of $V_1$ and $V_2$,
and  [up to the prefactor $\exp[h(r)]$] remains an exact eigenstate of the perturbed 
problem (with nonvanishing $V_1$ and $V_2$).
If this argument were universally applicable, then the Dirac equation should have at least
one exact solution for very wide classes
of potentials $V_0$, $V_1$ and $V_2$, which can be expressed
in closed analytic form [provided $f_0(r)$ and $g_0(r)$ admit such a form].
However, there is another relation that follows from
Eq.~\eqref{PartialRelation} through division, 
\begin{equation}
\label{restric1}
\frac{g(r)}{f(r)}=\frac{g_0(r)}{f_0(r)}
=\sqrt{\frac{V_1(r)+V_2(r)}{V_1(r)-V_2(r)}}\,,
\end{equation}
where we used Eq.~\eqref{PertORel} to relate the ratio of the original wave
functions to the ratio of the wave functions of the perturbed Hamiltonian.
This relation implies that the ``unperturbed''
upper and lower radial wave functions 
$f_0(r)$ and $G_0(r)$ must be mutually related by a
global prefactor. For convenience, we denote the 
proportionality factor as $-\gamma$, so that $f_0(r)=-\gamma\,g_0(r)$,
with reference to Sec.~\ref{sec2}.
Taking into account Eq.~\eqref{PertORel},
this also implies that $f(r)=-\gamma\,g(r)$.
We then find that fine-tuning of the coupling constants
of the two added potentials is required,
\begin{equation}
\label{restric2}
V_2(r)=-\frac{1-\gamma^2}{1+\gamma^2} \, V_1(r)\,.
\end{equation}
Since the value of $\gamma$ is determined by the relationship of the original
radial wave function, this turns out to be an extremely restrictive condition.
In the case where the original Hamiltonian is the Dirac--Coulomb Hamiltonian,
for example, the relationship between the two potentials is found to
be expressible in a simpler form (see Sec.~\ref{sec2})
\begin{equation}
\label{restric3}
\frac{1-\gamma^2}{1+\gamma^2} =\frac{E}{m} = 
\sqrt{1 - \frac{\lambda^2}{\kappa_0^2} } \,.
\end{equation}
For the classes of potentials discussed in Sec.~\ref{sec2},
and the states with $n_0 = -\kappa_0$,
all these relations~\eqref{restric1},~\eqref{restric2} and~\eqref{restric3} 
are fulfilled. Otherwise, the 
relations~\eqref{restric1},~\eqref{restric2} and~\eqref{restric3}
represent important restrictions which can be fulfilled, if at all,
only for particular solutions of the ``unperturbed'' problem (the one with 
$V_0$) and also limit the occurrence of the exact preservation to 
potentials $V_1$ and $V_2$ which have to be proportionally related 
according to Eq.~\eqref{restric2}.
Furthermore, as shown in Sec.~\ref{sec3}, the preservation
cannot be parametrically adjusted to more than
one reference state.

Yet, we are able to confirm that the energy eigenvalue preservation 
mechanism applies not only to {\em linear} confining potentials.
A valid application of the above considerations would concern 
the case
\begin{subequations}
\label{gen}
\begin{align}
V_0(r) =& \; -\frac{\lambda}{r} \,,
\qquad 
V_1(r) = A \, (r/r_0)^M \,, \qquad M = 1000 \,,
\\[2ex]
V_2(r)=& \; -\frac{1-\gamma^2}{1+\gamma^2} \, V_1(r) =
- \sqrt{1 - \frac{\lambda^2}{\kappa_0^2} } \, V_1(r) \,.
\end{align}
\end{subequations}
Here, the case $M=1000$ serves as a numerical approximation to 
a potential $V_1(r)$ which tends to zero for $r < r_0$, while it is 
numerically large for $r > r_0$, thus approximating the 
step-potential used in the construction of the MIT bag model.

%
%
\section{Conclusions}
\label{sec6}

Dirac equations with radially symmetric confining potential 
are of pheonomenological interest~\cite{HaJeSu1949,GM1949,GM1950I,GM1950II,
ChEtAl1974,MoSa1985,AbFu1987}.
Our considerations have been centered 
about a generalized Dirac Hamiltonian with two linear radial
confining potentials, as given in Eq.~\eqref{FrankHam}. Given the general
complexity of the Dirac equation, one would intuitively assume that the
addition of the confining potentials shifts the energy values of bound states,
no matter what the relation between the prefactors of the two confining
potentials in Eq.~\eqref{FrankHam} is, and irrespective of the quantum numbers
of the ``unperturbed'' Dirac--Coulomb eigenstates.  However, it can be
verified, both on the basis of an explicit solution of the radial Dirac
equation (see Sec.~\ref{sec5}) as well as on the basis of a perturbative calculation
(see Sec.~\ref{sec2}), that an interesting 
energy eigenvalue preservation occurs
provided the prefactor of the two confining potentials ($\mu$ and $\nu$), and
the ``Coulomb'' coupling parameter $\lambda$, fulfill the relationship given
in Eq.~\eqref{cancel}.  This preservation may occur for 
the reference $1S$ ground state
(if we set $n_0 = -\kappa_0 = 1$), which retains its 
precise energy eigenvalue under
the addition of the confining potentials,
but only under the condition that the ``fine-tuning relation''
given in Eq.~\eqref{allthestuff} is fulfilled.

As revealed by the FW transformed Hamiltonian given in
Eq.~\eqref{FWHamC}, the preservation is not fully accidental: Namely, the
leading-order effective operators [of order $\calO(\lambda^3\,m)$] derived from
the two confining potentials cancel for particle states [see
Eq.~\eqref{FWHamCplus}], while they add up for antiparticle states [see
Eq.~\eqref{HminusDCb}].  At order $\calO(\lambda^5\,m)$, the preservation is
more subtle and involves both the kinetic corrections in the confining
potential as well as the additional spin-orbit coupling terms which are
manifest in Eq.~\eqref{FWHamCplus}. Such spin-orbit terms are known to affect
bound Dirac particles in comparable situations, such as in nuclei where the
hadrons are confined by the short-range meson exchange potentials, and yet,
spin-orbit terms have a significant 
influence on the spectrum~\cite{MK1992,GrMa1996}.
The preservation is explicitly verified in Eq.~\eqref{R2}.

Yet, in Sec.~\ref{sec3}, we verify that the preservation can be tailored to
only one reference state $|n_0 \, \kappa_0\rangle$ with $\kappa_0 = - n_0$;
other states are shifted by the added potential. 
This result, expressed in Eq.~\eqref{ahahahaaaa}, would be impossible to obtain 
without the explicit Foldy--Wouthuysen transformed 
Hamiltonian being available (its matrix elements can be 
expressed analytically). Furthermore, the analysis of
corresponding antiparticle states, carried out in Sec.~\ref{sec4}, reveals that
the physical degrees of freedom are drastically different as compared to
particle states [for the nonrelativistic approximation
relevant to antiparticles, see Eq.~\eqref{HminusEFF}].

In summary, we hope to 
have accomplished three goals in this paper: {\em (i)}~First, to show that the 
considerations reported in Refs.~\cite{Fr1999,CaFr2000}
are restricted to one particular reference state of the
``double-confining'' 
potential given in Eq.~\eqref{FrankHam}. This restriction may
not be completely obvious from the papers~\cite{Fr1999,CaFr2000}
(see also~\ref{sec5}). 
Furthermore, while the preservation mechanism can be
tailored to {\em any} reference state with $n = n_0 = -\kappa_0$, 
it can be tailored to work for 
only one such state, not more than one (see Sec.~\ref{sec3}). Moreover, our
considerations show that the existence of the preservation mechanism
actually is tied to a certain proportionality of the upper and lower radial
wave functions of the ``unperturbed'' Dirac-Coulomb problem, which persists
only for $n = n_0 = -\kappa_0$. 
However, the preservation mechanism can be extended to 
potentials which model the step-like behavior of the MIT bag 
model~\cite{ChEtAl1974,MoSa1985}, as demonstrated in Eq.~\eqref{gen}.
{\em (ii)}~We identify, via a Foldy--Wouthuysen transformation, the
nonrelativistic effective operators which pertain to an intuitive physical
interpretation of the Dirac equation in the nonrelativistic limit. 
The calculation of the nonrelativistic operators reveals the 
existence of large spin-orbit coupling terms in the confining 
potentials and an interesting ``duality'' [see Eq.~\eqref{FWHamC}]:
The leading confining term cancels 
for a positive-energy particle state (but not for anti-particles),
while the spin-orbit term vanishes for antiparticle states 
(but not for positive-energy states). {\em (iii)}~Third, 
we identify the effective operators
that pertain to {\em antiparticles} as opposed to particles, in the double-confining
potential.  We show that the preservation mechanism
relevant for particles is reversed for the corresponding antiparticles which
are described by the same (generalized) Dirac equation. In passing, we obtain
the effective form of the confining potential for antiparticles and the
effective (fully confining) operators relevant in the nonrelativistic limit (for
antiparticles).  This latter conclusion illustrates the intricacies, and
perhaps, also the power, of the Dirac formalism: namely, to describe two
different particle ``species'' (particles and corresponding antiparticles) at
the same time, and with the same equation, predicting a different, and in our
case, opposite influence (addition versus preservation) of the two confining
potentials for the antiparticle as opposed to the particle states.

%
%
\section*{Acknowledgments}

The research has been supported by the National Science Foundation
(Grants PHY--1068547 and PHY--1403973).

\end{document}